# Passive Photoacoustic Effect

Yiyun Wang[#], Qingyuan Shi[#], Yuting Shen[#], Yifan Liu[*], and Fei Gao[*]

[1]Hybrid Imaging System Laboratory, School of Information Science and Technology, ShanghaiTech University, Shanghai 201210, China
[2]School of Physical Science and Technology, ShanghaiTech University, Shanghai 201210, China
[3]Chinese Academy of Sciences, Shanghai Institute of Microsystem and Information Technology, Shanghai 200050, China
[4]University of Chinese Academy of Sciences, Beijing 100049, China
[5]Shanghai Engineering Research Center of Energy Efficient and Custom AI IC, Shanghai 201210, China
[#] Equal contribution
[*] Corresponding author

Photoacoustic effect refers to the acoustic generation induced by laser irradiation, where nanosecond pulsed laser source is normally used to provide instantaneous heating and thermoelastic expansion of the sample. More generally, photoacoustic generation requires active intensity modulation of laser source to produce sharp temperature gradient, which is key to generate acoustic wave. In this paper, we propose a novel photoacoustic effect from moving droplet, which generates photoacoustic wave by a non-modulated continuous-wave laser. When the droplet moves through the laser spot area, it can be heated up and cooled down instantaneously, generating photoacoustic waves. We name it passive photoacoustic effect. Theoretical analysis and simulation study validated the existence of passive photoacoustic effect. This phenomenon may find potential application in high-throughput photoacoustic cytometry.

*Introduction*.

Since discovered by Alexander Graham Bell in 1880s [1,2], photoacoustic (PA) effect (also called optoacoustic effect) has been studied and applied in many applications, such as gas sensing [3,4], biomedical imaging [5,6], flow cytometry [7–9], etc. To generate PA effect, pulsed or intensity-modulated laser source is utilized to provide instantaneous heating and sharp temperature gradient in the sample [10–15]. In other words, the intensity of laser fluence needs to be actively modulated. Wenyu et al. recently reported the analytical modelling of PA sources by moving a CW laser source [16,17], which still needs active moving of the laser spot at extremely high speed that is close to sound speed.

Is it possible to generate PA signal by a stationary non-modulated continuous-wave (CW) laser source? In this paper, we explore a novel type of the PA effect, and define it as passive PA effect. It describes the passive excitation mechanism of photoacoustic signal generation, which is induced by the flowing of droplet. The conventional excitation of PA effect is mainly by laser intensity modulation, actively. Instead, passive PA effect generates acoustic wave from moving droplet, illuminated by a stationary non-modulated CW laser. When it moves into a CW laser spot, a droplet is heated up and generates PA wave due to thermoelastic expansion. Similarly, owing to the instantaneous contraction, the droplet is cooled down and generates passive PA wave when it leaves the laser spot. In this paper, we theoretically analyze the existence of passive PA phenomenon and perform simulation validation. We also discuss some key parameters in designing a passive photoacoustic flowmetry.

*Theory*.

To build a general model and analyze the phenomenon, we make the following assumptions. The droplet's moving speed is fast enough to satisfy the thermal confinement for PA signal generation. The CW laser energy is within safety limit, so that the droplet is not blasted during laser radiation period. We also assume that the droplet region in the laser spot accumulates heat with negligible heat dissipation and diffusion. When the partial slice of droplet leaves laser spot, its accumulated heat is dissipated instantaneously. The passive PA effect mainly has two kinds of setups: laser spot diameter is smaller than the length of droplet (Case 1 shown in Fig. 1); laser spot diameter is larger than the length of droplet (Case 2 shown in Fig. 2). We will further analyze two cases, respectively.

In Case 1, the laser spot diameter is smaller than the length of droplet (Fig. 1). During the transition period T1-T3, the droplet moves rightwards, meeting the laser spot gradually. Heat begins to accumulate at the right edge of the droplet, leading to thermal expansion and positive PA signal generation. When the laser spot fully illuminates on the droplet, the temperature rising at the right edge of the droplet reaches maximum, as well as positive PA signal. Supposing the laser spot has diameter of $d$ and constant power of $W$ with homogeneous fluence distribution, the equivalent laser energy

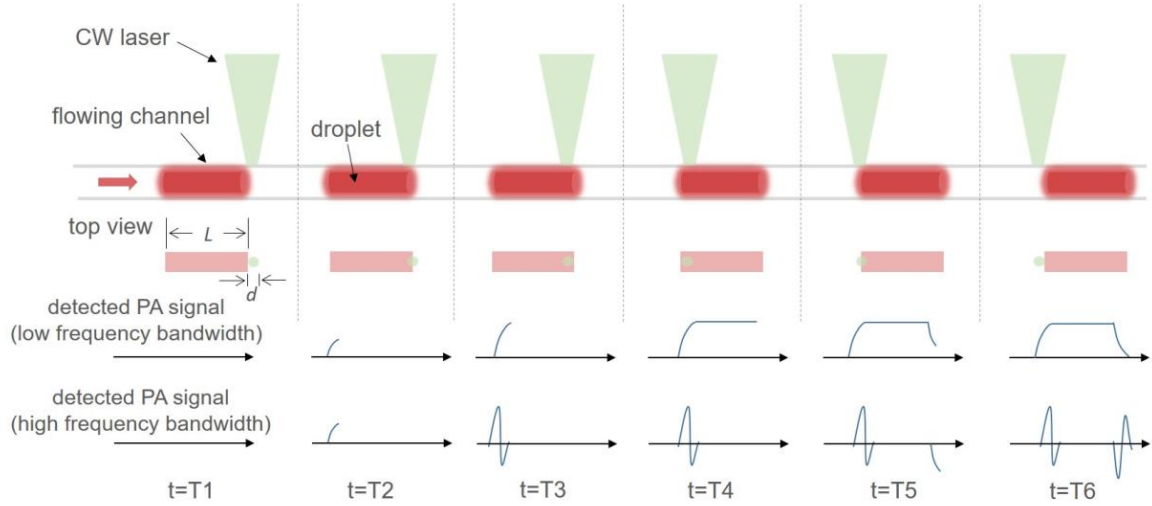

FIG. 1. The principle of passive PA effect induced by a flowing droplet (Case 1).

density $F$ during the positive transition period (T1~T3) can be approximately calculated by:

$$F_x = \frac{\bar{E}}{\bar{A}} = \frac{\frac{1x}{2v}\frac{\pi\left(\frac{x}{2}\right)^2}{\pi\left(\frac{d}{2}\right)^2}W}{\frac{1}{2}\pi\left(\frac{x}{2}\right)^2} = \frac{4xW}{\pi v d^2} \tag{1}$$

where $\bar{E}$ is the average energy in the droplet, $\bar{A}$ represents the average irradiated region, and constant $v$ is the moving velocity of droplet. Variable $x$ is the droplet's position. Since the laser spot is gradually moving onto the droplet, both the total energy $\frac{x}{v}\frac{\pi\left(\frac{x}{2}\right)^2}{\pi\left(\frac{d}{2}\right)^2}W$ and total spot area $\pi\left(\frac{x}{2}\right)^2$ are halved. Then the PA pressure $p_{x,0}$ at the initial laser excitation in position $x$ could be estimated by substituting Eq. (1) into the conventional PA generation equation [18], leading to:

$$p_{x,0} = \Gamma_0 \eta_{th} \mu_a F_x = \Gamma_0 \eta_{th} \mu_a \frac{4xW}{\pi v d^2} \tag{2}$$

where $\Gamma_0$ is the Grueneisen parameter at baseline temperature, $\eta_{th}$ is the heat conversion efficiency, $\mu_a$ is the optical absorption coefficient. During period T1~T3, there also exists heat accumulation in the irradiated region. According to the Grueneisen relaxation effect, the total PA pressure $p_x$ can be expressed as:

$$p_x = (\Gamma_0 + \Delta\Gamma)\eta_{th}\mu_a F_x \tag{3}$$

where $\Delta\Gamma$ is the Gruneisen coefficient increment due to temperature rising during period T1~T3, which can be expressed as [18]:

$$\Delta\Gamma = b\eta_{th}\mu_a F_n = b\eta_{th}\mu_a \int_0^x F_i di$$
$$= b\eta_{th}\mu_a \frac{2W}{\pi v d^2} x^2 \tag{4}$$

where $F_n$ is the accumulated energy density deposited during the period T1~T3. $b$ is the coefficient relating the thermal energy absorbed to the Grueneisen coefficient change. Substitute Eq. (4) into Eq. (3), we have:

$$p_x = \left(\Gamma_0 + b\eta_{th}\mu_a \frac{2W}{\pi v d^2} x^2\right)\eta_{th}\mu_a \frac{4xW}{\pi v d^2}$$

$$= p_{x,0} + b\eta_{th}^2\mu_a^2 \frac{8W^2 x^3}{\pi^2 v^2 d^4} \tag{5}$$

Thus, we can get the total PA pressure at t = T3 when $x = d$:

$$\begin{aligned} p_d &= p_{d,0} + b\eta_{th}^2\mu_a^2 \frac{8W^2}{\pi^2 v^2 d} \\ &= \Gamma_0\eta_{th}\mu_a \frac{4W}{\pi v d} + b\eta_{th}^2\mu_a^2 \frac{8W^2}{\pi^2 v^2 d} \end{aligned} \tag{6}$$

When the laser spot continues heating slices of the droplet in the transition period T3-T4, the area under laser spot keeps the light energy deposit and temperature elevation. It leads to constant thermal expansion pressure $p_x$. As the droplet gradually moves out of the laser spot (T4-T6), heat begins to decline at the right edge of the droplet, leading to contraction and decreased PA pressure till zero.

From Eq. (5), we could observe that the amplitude of PA signal is inversely proportional to the moving speed of droplet, and diameter of the laser spot. It is reasonable because a lower speed leads to longer transition time, i.e. longer laser illumination time and more energy deposit. From the expected PA waveforms shown in Fig. 1, we could see that there will be a main center frequency, which can be estimated as:

$$f_l = \left(2 \times \frac{d+L}{v}\right)^{-1} \tag{7}$$

It can be seen that the central frequency is proportional to the moving speed of droplet, and inversely proportional to the size of droplet. It is also inversely proportional to the length of the droplet. We also expect that there exists high-frequency PA signal when the heat accumulation in the droplet has transient variation.

In Case 2, we discuss the situation that laser spot diameter is larger than the length of droplet (Fig. 2). Similar with Case 1, we may reach PA pressure $p_L$ by replacing $x$ with $L$ in Eq. (5), at the positive transition moment T3:

$$p_L = \Gamma_0\eta_{th}\mu_a \frac{4WL}{\pi v d^2} + b\eta_{th}^2\mu_a^2 \frac{8W^2 L^3}{\pi^2 v^2 d^4} \tag{8}$$

In the period from T3 to T4, the droplet continues to move rightwards and accumulates heat due to light illumination. Temperature of the droplet will be increased leading to larger PA pressure $p_2$ at t = T4, based on Grueneisen relaxation effect.

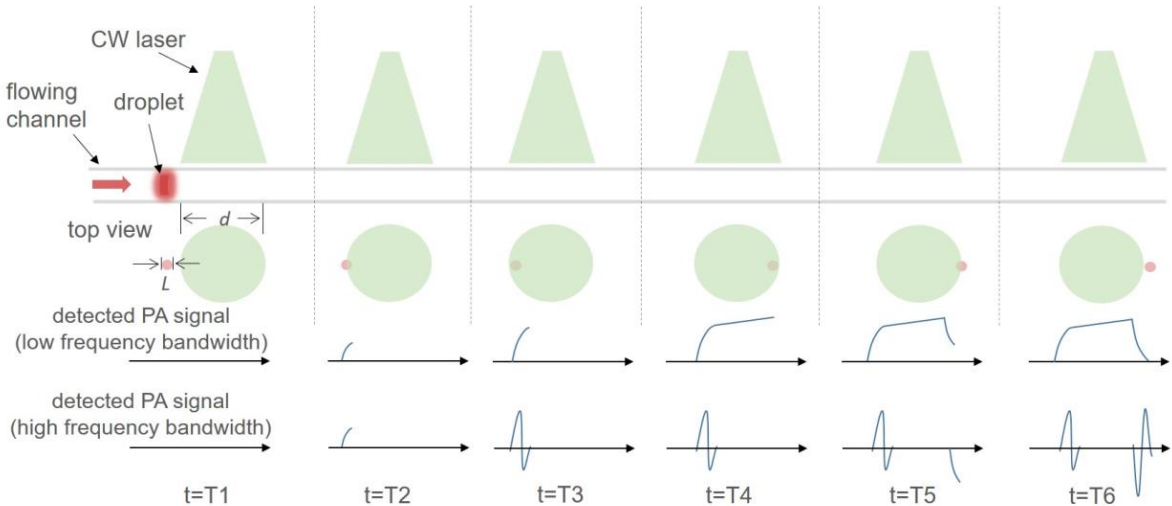

FIG. 2. The principle of passive PA effect induced by a flowing droplet (Case 2).

$$p_d = (\Gamma_0 + \Delta\Gamma)\eta_{th}\mu_a \frac{4WL}{\pi v d^2} \quad (9)$$

where $\Delta\Gamma$ can be expressed as:

$$\Delta\Gamma = b\eta_{th}\mu_a \int_0^d F_i di = b\eta_{th}\mu_a \frac{2W}{\pi v} \quad (10)$$

where $F_i$ is the accumulated energy density deposited at position $i$ in period T3~T4. Substitute Eq. (10) into Eq. (9), we have:

$$p_d = \Gamma_0 \eta_{th}\mu_a \frac{4WL}{\pi v d^2} + b\eta_{th}^2\mu_a^2 \frac{8W^2 L}{\pi^2 v^2 d} \quad (11)$$

During the transition period T4-T6, similar with the first case, the enhanced PA pressure descends until it reaches minimum.

From Eq. (11), we could also observe that the enhancement of PA pressure is proportional to the optical absorption coefficient, length of the droplet and laser power. On the other hand, it is inversely proportional to the moving speed and laser spot diameter.

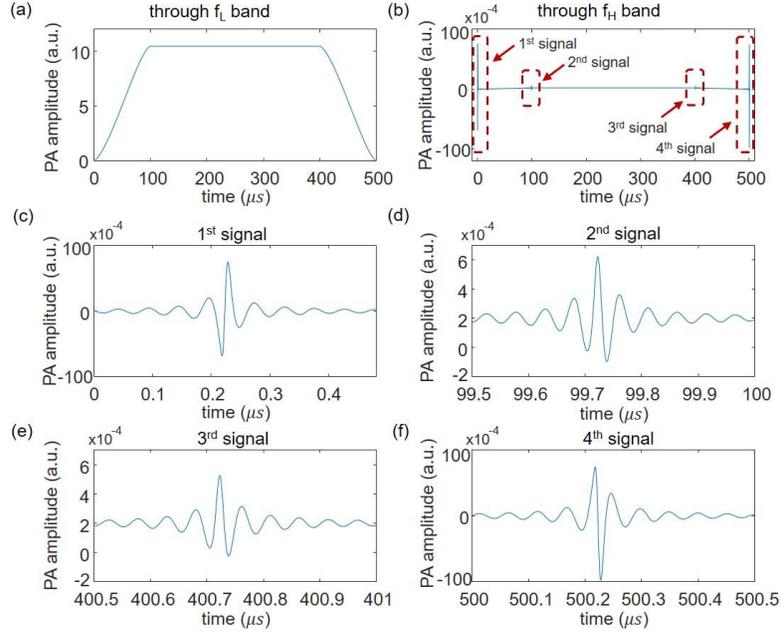

FIG. 3. The passive PA simulation results under Case 1. (a) PA signal's low-frequency component and (b) high-frequency component. (c)-(f). Zoom-in waveforms of the four PA signals in (b).

*Simulation study.*

Here, we simulate the passive PA effect using MATLAB k-Wave toolbox. It utilizes the k-space pseudo-spectral time-domain solution for PA generation and propagation simulation[19]. We first validate the phenomenon in two cases. Then, the influence of variables on passive PA generation is investigated. At the aim of better simulating practical scenarios, we place the single-point ultrasound sensor in front of the laser spot's center, and the sensor collects the simulated one-way passive PA wave propagation with attenuation.

In Case 1, we set the diameter of laser spot as 100 μm, and the droplet length as 400 μm. The droplet's moving velocity is set as 1 m/s. We obtain the PA signal's low-frequency component through a 1 kHz low-pass filter as predicted. According to the simulation results in Fig. 3(a), the low-frequency component ascends with the increasing heat accumulation and irradiated region, and vice versa. When the laser spot fully irradiates on the moving droplet, the heat within the droplet remains the same and the droplet system is in equilibrium state, which can be observed in the middle stage of Fig. 3(a). Fig. 3(b) displays the transient thermal variations during the process by applying band-pass filtering. By observing Fig. 3(c) and 3(f), we find that PA signals exhibit opposite phases, which represent the transient variation of thermal expansion and contraction, respectively. These two signals have much larger amplitudes than those in Fig. 3(d) and 3(e), reflecting sharper transient changes of thermal accumulation at T1 and T6. As shown, the PA signal's amplitude in Fig. 3(e) is slightly smaller than that in Fig. 3(d). It means that the temperature variation at T5 is milder than that at T3, since part of the droplet starts to leave the laser spot at T5. The simulation results fit the aforementioned theoretical analysis of passive PA effect well for Case 1.

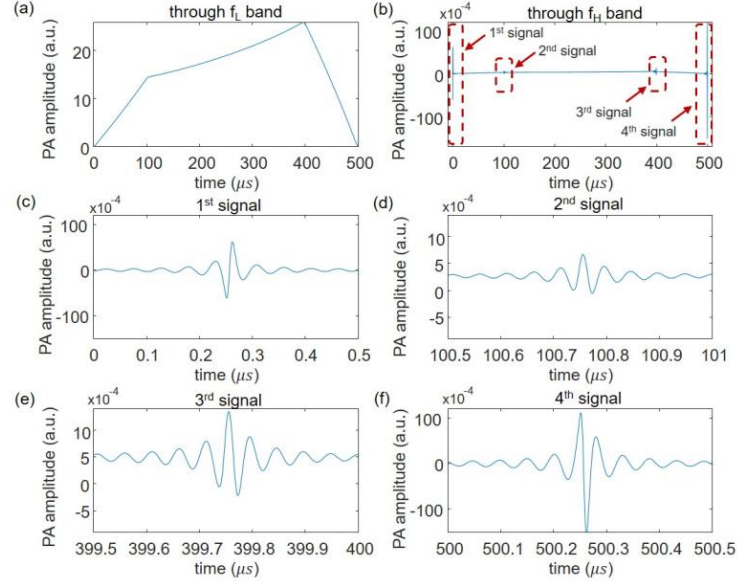

FIG. 4. The passive PA simulation results under Case 2. (a) PA signal's low-frequency component and (b) high-frequency component. (c)-(f). Zoom-in waveforms of the four PA signals in (b).

We then validate Case 2 by setting the droplet length as 100 μm and laser spot's diameter as 400 μm. The period T1-T3 and T4-T6 are similar to those in Case 1. During the transition period T3-T4, the entire droplet is irradiated, and heat is continuously accumulating, which reflects as an ascending curve in the PA signal (Fig. 4(a)). In Case 2, the signal's amplitude at $t = $ T4 (Fig. 4(e)) is larger than that at $t = $ T3 (Fig. 4(d)), since the droplet accumulates far more heat at $t = $ T4. From Fig. 4(c) and 4(f), we also observe PA signals with opposite phases. It is also noticed that the amplitude of PA signal in Fig. 4(f) is larger than that in Fig. 4(c), which is owing to the heat accumulation.

Through comparing results by varying key parameters, we are able to observe their influences on the passive PA generation. In Case 1, we observe that the droplet's moving speed $v$ and the laser spot's diameter $d$ are two key parameters. As deduced in the theoretical analysis, the amplitude of PA signal is inversely proportional to the laser spot's diameter. The simulation results are shown in Fig. 5(a). In the simulation, we remain the laser energy as a constant. It leads to a larger maximum amplitude of PA signal when the droplet is irradiated by a laser spot with 70 μm diameter. In another simulation, we vary the droplet's flowing velocities. Shown in Fig. 5(b), the amplitude of the signal ascends with the descending velocity. For Case 2, we select droplet's moving speed $v$ and the length of droplet $L$ as key parameters, and discuss their influences to the simulation results. Illustrated in Fig. 6(a) and 6(b), PA signal's amplitude is inversely proportional to the droplet's velocity, and is proportional to the length of droplet, same with the theoretical deduction in Case 2 as expected.

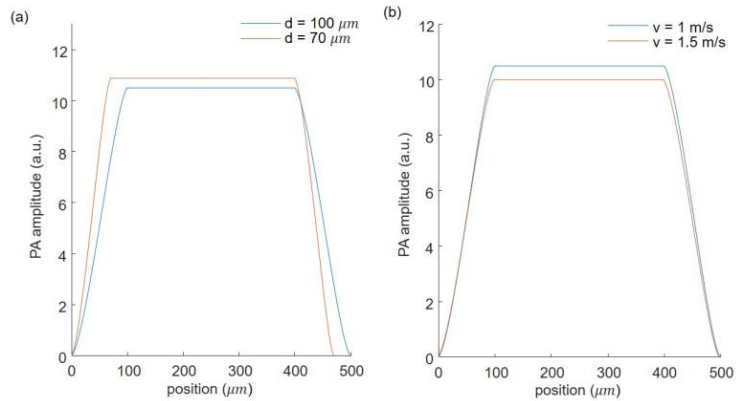

FIG. 5. The passive PA signal results under Case 1 with various key parameters. (a). PA signal with different laser spot's diameters. (b). PA signal with different droplet's flowing speeds.

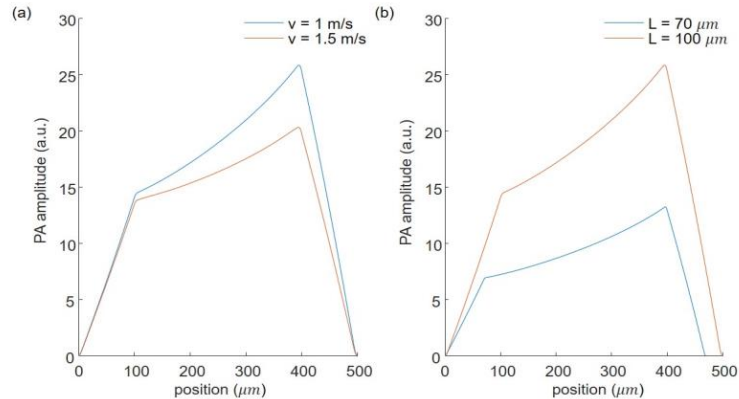

FIG. 6. The passive PA signal results under Case 2 with various key parameters. (a). PA signal with different droplet's flowing speed. (b). PA signal with different droplet lengths.

*Conclusion*.

In the paper, we propose a new type of PA generation mechanism, passive PA effect. This phenomenon is based on stationary continuous-wave laser illumination on a moving droplet. We investigate several key parameters to the generation of passive PA effect, and demonstrate this phenomenon in two cases through theoretical analysis and simulation validation. The passive PA effect may find potential application in high-throughput PA flowmetry.

[1] G. Bell, Alexander, *The Production of Sound by Radiant Energy*, Science (1979) **2**, 242 (1881).
[2] G. Bell, Alenxander, *On the Production and Reproduction of Sound by Light*, American Journal of Science **s3-20**, 305 (1880).
[3] M. Nägele and M. W. Sigrist, *Mobile Laser Spectrometer with Novel Resonant Multipass Photoacoustic Cell for Trace-Gas Sensing*, Applied Physics B: Lasers and Optics **70**, 895 (2000).
[4] S. Manohar and D. Razansky, *Photoacoustics: A Historical Review*, Advances in Optics and Photonics **8**, 586 (2016).
[5] L. v. Wang and S. Hu, *Photoacoustic Tomography: In Vivo Imaging from Organelles to Organs*, Science (1979) **335**, 1458 (2012).
[6] A. Taruttis and V. Ntziachristos, *Advances in Real-Time Multispectral Optoacoustic Imaging and Its Applications*, Nature Photonics **9**, 219 (2015).
[7] V. P. Zharov, E. I. Galanzha, E. v. Shashkov, N. G. Khlebtsov, and V. v. Tuchin, *In Vivo Photoacoustic Flow Cytometry for Monitoring of Circulating Single Cancer Cells and Contrast Agents*, Optics Letters **31**, 3623 (2006).
[8] E. I. Galanzha et al., *In Vivo Liquid Biopsy Using Cytophone Platform for Photoacoustic Detection of Circulating Tumor Cells in Patients with Melanoma*, Science Translational Medicine **11**, (2019).
[9] D. A. Nedosekin, T. Fahmi, Z. A. Nima, J. Nolan, C. Cai, M. Sarimollaoglu, E. Dervishi, A. Basnakian, A. S. Biris, and V. P. Zharov, *Photoacoustic in Vitro Flow Cytometry for Nanomaterial Research*, Photoacoustics **6**, 16 (2017).
[10] S. Palzer, *Photoacoustic-Based Gas Sensing: A Review*, Sensors (Switzerland) **20**, (2020).
[11] A. B. E. Attia, G. Balasundaram, M. Moothanchery, U. S. Dinish, R. Bi, V. Ntziachristos, and M. Olivo, *A Review of Clinical Photoacoustic Imaging: Current and Future Trends*, Photoacoustics **16**, 100144 (2019).
[12] L. v. Wang, *Tutorial on Photoacoustic Microscopy and Computed Tomography*, IEEE Journal on Selected Topics in Quantum Electronics **14**, 171 (2008).
[13] S. Kellnberger, D. Soliman, G. J. Tserevelakis, M. Seeger, H. Yang, A. Karlas, L. Prade, M. Omar, and V. Ntziachristos, *Optoacoustic Microscopy at Multiple Discrete Frequencies*, Light: Science and Applications **7**, (2018).
[14] R. Zhang, F. Gao, X. Feng, S. Liu, R. Ding, and Y. Zheng, *Photoacoustic Resonance Imaging*, IEEE Journal of Selected Topics in Quantum Electronics **25**, 1 (2019).
[15] H. Fang, K. Maslov, and L. v. Wang, *Photoacoustic Doppler Effect from Flowing Small Light-Absorbing Particles*, Physical Review Letters **99**, 2 (2007).
[16] W. Bai and G. J. Diebold, *Photoacoustic Effect Generated by Moving Optical Sources: Motion in One Dimension*, Journal of Applied Physics **119**, (2016).
[17] W. Bai and G. J. Diebold, *Moving Photoacoustic Sources: Acoustic Waveforms in One, Two, and Three Dimensions and Application to Trace Gas Detection*, Journal of Applied Physics **125**, (2019).

[18] L. Wang, C. Zhang, and L. v. Wang, *Grueneisen Relaxation Photoacoustic Microscopy*, Physical Review Letters **113**, 1 (2014).
[19] B. E. Treeby and B. T. Cox, *K-Wave: MATLAB Toolbox for the Simulation and Reconstruction of Photoacoustic Wave Fields*, Journal of Biomedical Optics **15**, 021314 (2010).